\documentclass[preprint,12pt]{elsarticle}

\usepackage{textcomp}
\usepackage{color}
\usepackage{amssymb,amsmath}
\usepackage{mathrsfs}
\usepackage{float}
\usepackage{graphicx}

\biboptions{sort,compress}

\journal{Nuclear Instruments and Methods A}
\begin{document}
\begin{frontmatter}
\title{TITAN's Digital RFQ Ion Beam Cooler and Buncher, Operation and Performance}
\author[label1,label2]{T.~Brunner\corref{cor1}}
\cortext[cor1]{Corresponding Author}
\ead{thomas.brunner@triumf.ca}
\author[label1,label3]{M.J.~Smith}
\author[label1,label3]{M.~Brodeur\fnref{fn1}}
\author[label1,label3]{S.~Ettenauer}
\author[label1,label3]{A.T.~Gallant}
\author[label1,heidelberg,label4]{V.V.~Simon}
\author[label1]{A.~Chaudhuri}
\author[label1]{A.~Lapierre\fnref{fn1}}
\author[label1]{E.~Man\'e}
\author[label1]{R.~Ringle\fnref{fn1}}
\author[label1]{M.C.~Simon}
\author[label1]{J.A.~Vaz}
\author[label3]{P.~Delheij}
\author[label1]{M.~Good}
\author[label1]{M.R.~Pearson}
\author[label1,label3]{and J.~Dilling}
\address[label1]{TRIUMF, 4004 Wesbrook Mall, Vancouver, BC, V6T 2A3, Canada}
\address[label2]{Physik Department E12, Technische Universit\"at M\"unchen, James Franck Stra\ss e, D-85748 Garching, Germany}
\address[label3]{Department of Physics and Astronomy, University of British Columbia, 6224~Agriculture Road, Vancouver, BC, V6T 1Z1, Canada}
\address[label4]{Max-Planck-Institut f\"ur Kernphysik, Saupfercheckweg 1, Heidelberg, Germany}
\address[heidelberg]{University of Heidelberg, D-69117 Heidelberg, Germany}
\fntext[fn1]{Current address: NSCL, Michigan State University, East Lansing, MI, 48824, USA}

\begin{abstract}
We present a description of the Radio Frequency Quadrupole (RFQ) ion trap built as part of the TITAN facility. It consists of a gas-filled, segmented, linear Paul trap and is the first stage of the TITAN setup with the purpose of cooling and bunching radioactive ion beams delivered from ISAC-TRIUMF. This is the first such device to be driven digitally, i.e., using a high voltage ($V_{pp} = \rm{400 \, V}$), wide bandwidth ($0.2 < f < 1.2 \, \rm{MHz}$) square-wave as compared to the typical sinusoidal wave form. Results from the commissioning of the device as well as systematic studies with stable and radioactive ions are presented including efficiency measurements with stable $^{133}$Cs and radioactive $^{124, 126}$Cs. A novel and unique mode of operation of this device is also demonstrated where the cooled ion bunches are extracted in reverse mode, i.e., in the same direction as previously injected.
\end{abstract}

\end{frontmatter}
\section{Introduction}
\label{intro}
The use of ion traps has been a standard technique to manipulate ions for many years. One typical application is the use of linear ion traps in mass spectrometry~\cite{Dou05}. But also at radioactive ion beam facilities ion traps are applied as a standard technique for the production of bunched ions with low emittance~\cite{Bla06}. The gas-filled, radio-frequency quadrupole (RFQ) is such a device \cite{Kel01} whereby a previously decelerated, low energy  ($E<100\, \rm{eV}$) beam is thermalized via successive collisions with an inert buffer gas. Left unchecked, this cooling process would lead to radial dispersion of the ion beam and eventually to its loss. However, by application of an oscillating electric potential (typically $\rm{V}_{\rm{pp}} \approx \, \rm{100 \, V}$, $f \approx \, \rm{1 \, MHz}$) to a quadrupolar electrode structure, a net radial force can be created which acts to counter this dispersion allowing ion beams to be cooled quickly and efficiently.

The first RFQ cooler was built as part of the ISOLTRAP~\cite{Muk08,Bol96} experiment in 1998 and was designed to match the properties of the incoming ISOLDE~\cite{Kug00} beam to the acceptance of their Penning trap mass spectrometer~\cite{Her01,Kel01}. Following the success of ISOLTRAP, a number of new Penning trap facilities have been built worldwide in order to exploit the availability of different radioactive isotopes at other on-line facilities such as JYFLTRAP at IGISOL~\cite{Kolhinen:2004la}, SHIPTRAP at GSI~\cite{Block:2005it}, CPT at ANL~\cite{Savard:1997ly}, and LEBIT at NSCL~\cite{0953-4075-36-5-313}. An RFQ cooler is a common feature of all these experiments and is now recognized as a vital component for the preparation of ions for high-precision mass measurements. More recently, they have been used as a general tool for beam preparation~\cite{Pod04,Fra08}. Moreover, other uses for cooled ion pulses have also been found, most notably in the field of collinear laser spectroscopy, whereby beam cooling greatly reduces the Doppler broadening of the observed spectral lines and bunched beams improve the signal to background ratio by orders of magnitude~\cite{PhysRevLett.88.094801}. The use of high ($\rm{V}_{\rm{pp}} > 1 \rm{kV}$) RF driving fields has also been investigated, a requirement for the trapping and cooling of high intensity ion beams\cite{Moore:2006lr}. 

In this paper we present results from the commissioning and operation of the RFQ cooler and buncher, which has been built as part of the TITAN (T{\sc riumf}'s Ion Traps for Atomic and Nuclear science) experimental facility~\cite{Dilling2006198,Delheij:2006ht} located at TRIUMF-ISAC~\cite{Dom00}. TITAN's science program includes precision mass measurements and laser spectroscopy~\cite{Man11} as well as in-trap decay spectroscopy measurements for double beta decay experiments~\cite{Bru11}. During TITAN's Penning trap mass measurements of halo nuclei $^8$He~\cite{Ryj08}, $^{11}$Li~\cite{Smi08} and $^{11}$Be~\cite{Rin09} the RFQ has been operated successfully, as well as during the mass measurements of radioactive $^{9,10,12}$Be~\cite{Rin09,Ett10} and neutron-rich Ca and K isotopes~\cite{Lap10}. 
Although essentially working on the same principle as all other RFQ coolers in operation, the TITAN RFQ possesses a number of unique features. First, unlike other traps which are driven sinusoidally, the TITAN RFQ is operated using a wide bandwidth ($200\,\rm{kHz} < f < 1.2 \, \rm{MHz}$) square-wave driver. Second, the RFQ is designed such that ion pulses can be extracted from either end of the cooler, allowing for operation in a reverse extraction mode whereby the cooled ions can be extracted back into the ISAC beam line from which the ions were initially injected. This technique has been applied with success to cool and bunch stable and short-lived Rb isotopes for collinear laser-spectroscopy studies~\cite{Man10,Man11}. Finally, the RFQ uses a unique set of harmonic deceleration optics which poses a solution to the well known problem of injecting a strongly decelerated ion beam into the RFQ beam cooler~\cite{Moore:2006lr}.
\section{Beam Cooling in a Digital RFQ}
\label{theory}
\subsection{Beam Cooling Basics}
\begin{figure}
\centering
\includegraphics[width=1\textwidth]{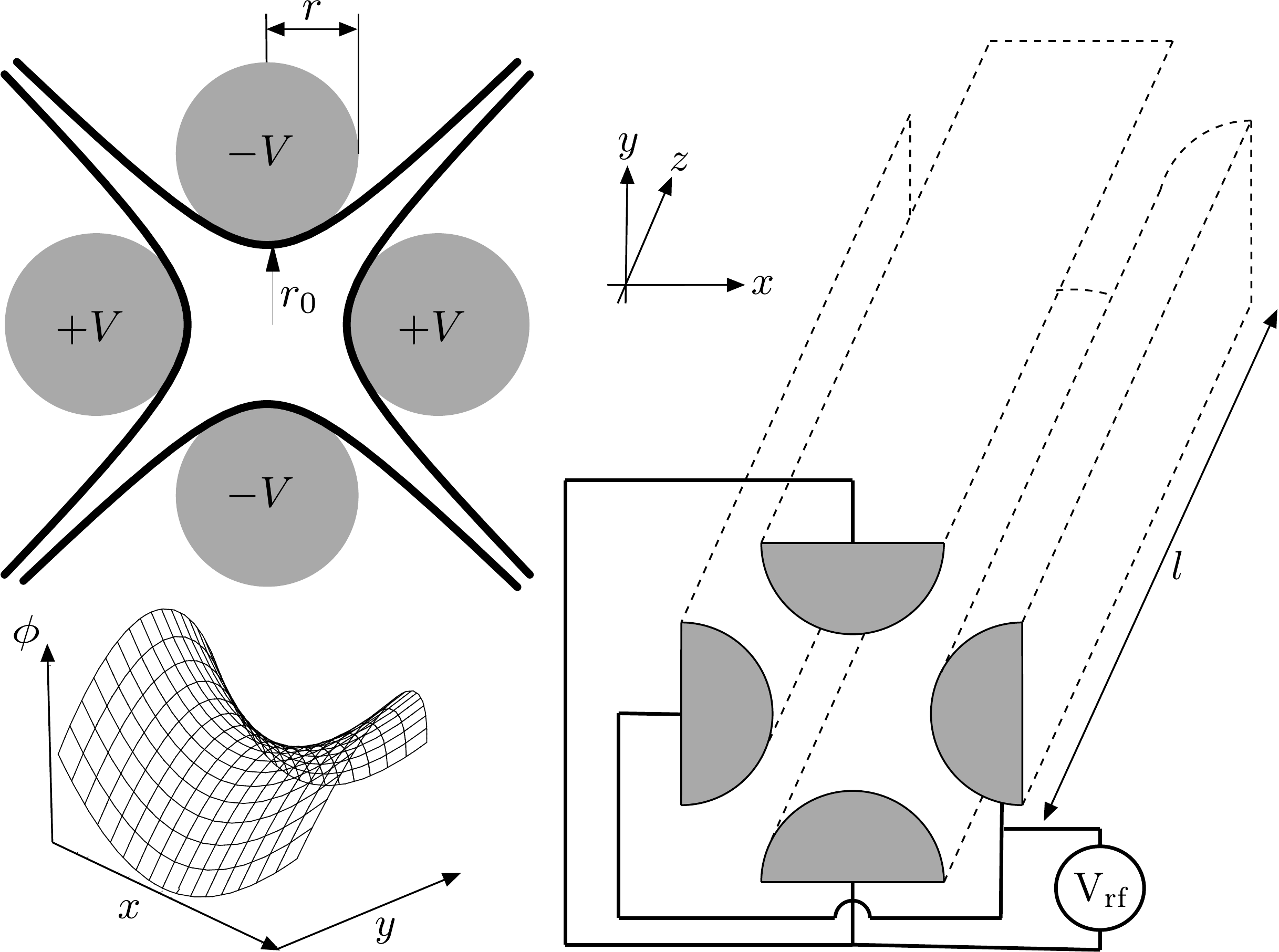}
 \caption{A schematic of an ideal hyperbolic electrostatic quadrupole (thick line) and the circular electrodes used to closely approximate the hyperbolic shape are shown on the top left. The saddle-shaped harmonic potential it creates is illustrated at the bottom left. The geometry of an RFQ formed from half-cylindrical rods and pair-wise applied alternating potential (V$_{\rm{rf}}$) is shown on the right. 
 \label{fig:rfq}}
\end{figure}

In an RFQ beam cooler, the temperature of an ion beam is reduced via successive collisions with the atoms of a buffer gas~\cite{Daw95,Sch08a}. By applying an RF field to a quadrupole electrode structure, one can provide a radial force which counteracts the dispersion of the ions caused by the cooling process. A standard electrostatic quadrupolar electrode structure, shown in Figure~\ref{fig:rfq}, creates a harmonic saddle-shaped potential which focuses ions linearly in one direction whilst defocussing in the perpendicular direction. However, by placing a series of quadrupolar devices together such that the focusing direction of each subsequent section is perpendicular to the previous, a net focusing force can be obtained~\cite{Courant:1958fj}. Alternatively, the same effect can be obtained by periodically switching the polarity of the voltage applied to the electrodes, thus forming a linear Paul trap. The radial motion of ions in such a trap is, to first order, harmonic.

\begin{figure}
\centering
\includegraphics[width=1\textwidth]{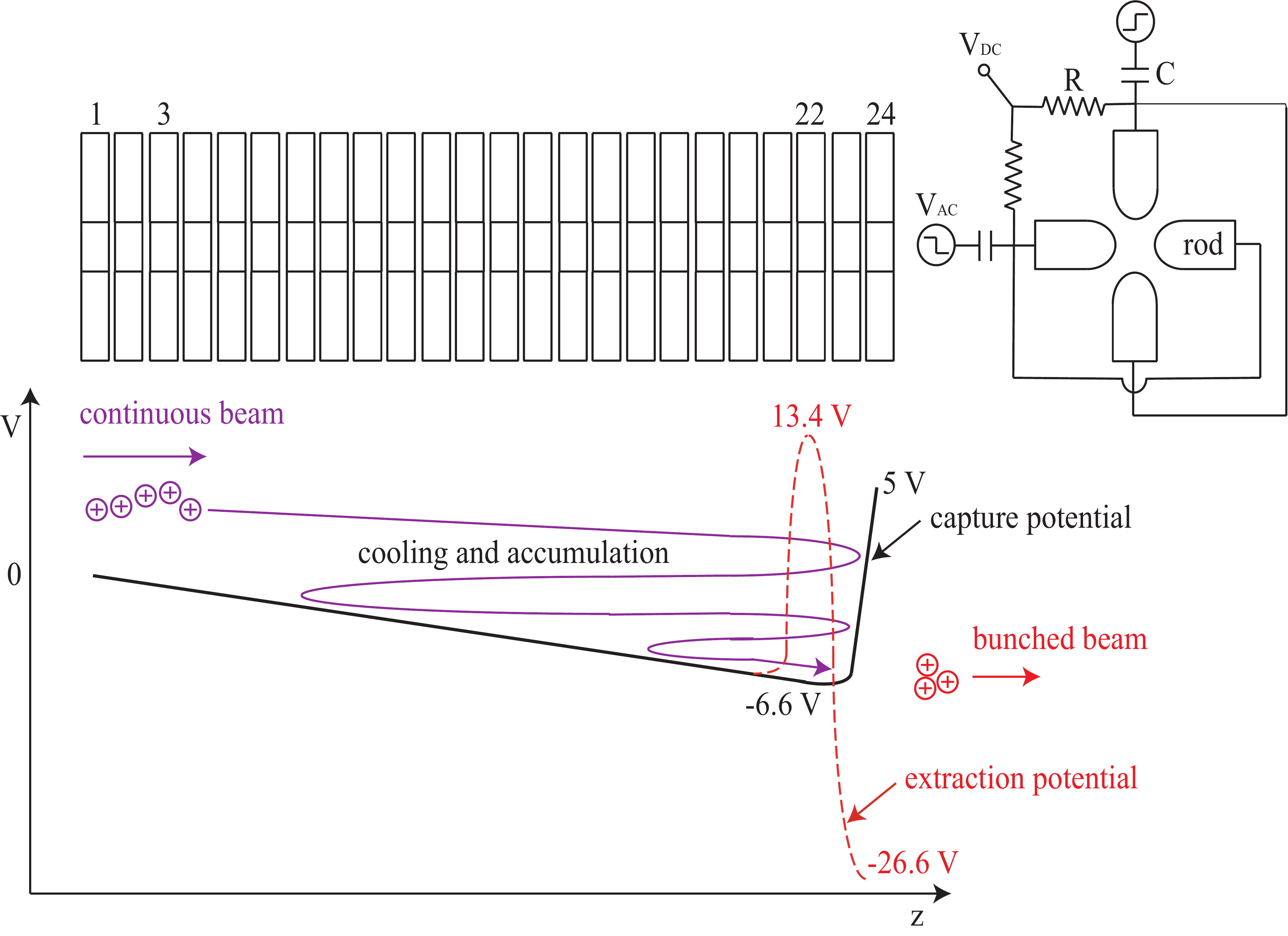}
 \caption{Schematic of the segmented electrodes of the TITAN-RFQ (top left) as well as the corresponding longitudinal drag potential typically applied to the electrodes (bottom left). A schematic of the applied voltages to the quadrupole electrodes is displayed in the top right. For explanation see text. \label{fig:rfq-potential}}
\end{figure}

A typical RFQ beam cooler for short-lived isotopes from radioactive on-line facilities is formed by placing four long circular rods, typically $l \approx 1 \, \rm{m}$, in a quadrupolar configuration as shown in Figure~\ref{fig:rfq}. The cooler is filled with an inert buffer gas, typically helium, at a pressure on the order of $1 \times 10^{-2} \, \rm{mbar}$. Ions are injected into the cooler such that the ion beam is centered on its longitudinal axis. The radial ion motion is then damped by the buffer gas and the ions are cooled to the temperature of the gas. A longitudinal electrostatic field gradient can be generated by segmenting the RFQ electrodes and applying differential potentials or adding electrostatic gradient rods~\cite{Bol04,Sun05}. In either case the field gradient inside the trap guides the ions through the device. By creating a potential minimum in the longitudinal direction, the ions are accumulated once the cooling process commences. A cooled bunch of ions is formed and can be extracted by pulsing the electrostatic potential applied to the electrode segments. A typical drag potential applied at TITAN's RFQ is displayed in Figure~\ref{fig:rfq-potential}. Ions are injected from the left and cooled in collisions with the buffer gas. The applied potential (solid line in Figure~\ref{fig:rfq-potential}) captures the ions at the potential minimum at electrode 23. In order to extract an ion bunch, the so-called guard electrode 24 is switched to a lower potential while the previous electrode 22 is switched to a higher potential in order to `kick' the ions out of the RFQ (dashed line in Figure~\ref{fig:rfq-potential}). 

\subsection{Motivation for a Digital RFQ}
Traditionally, RFQ ion cooler devices are driven with a
sinusoidally-varying RF potential. Ferrite cores are then used to split the
RF phase before it is applied to the trap electrodes. The leakage inductance of the transformer and the capacitance of the RFQ electrodes form an LC circuit on the secondary of the transformer~\cite{Smi05}.

The stability of an ion's motion in an RFQ is dependent both on the amplitude and frequency of the applied potential, as well
as on the charge-to-mass ratio $q/m$ of the ion and the geometry of the trap (see for example~\cite{Gho95}). It is desirable to keep the amplitude of the RF as large as possible as this increases the depth of the trapping potential and hence the transfer efficiency of the trap~\cite{Deh67,Mar05}. This is particularly important
for low-intensity radioactive ion beams. In a traditional RFQ driver, the transformer is operated far from the resonant frequency of the LC circuit such that the frequency response is flat in the frequency region of interest. This leads to a large amount of power dissipation in the secondary circuit and is typically not efficient. The latter design cannot be used at higher voltages as the power requirements are too severe. Instead, the transformer must be run close to the resonant frequency to reduce losses in the secondary circuit. Losses can be further reduced by using an LC circuit with a high quality factor $Q$, i.e., low bandwidth, resonator. This implies that a combination of high voltage and high bandwidth is impossible with a traditional driver.

The feasibility of a square-wave-driven, linear Paul trap was first demonstrated by J.~A.~Richards~\cite{Richards:1973lr}. Using bipolar junction transistors, a square-wave of $80 \, \textrm{V}$ maximum peak-to-peak amplitude was generated at frequencies of up to $1 \, \textrm{MHz}$. This was used to drive a small quadrupole mass filter. The use of square-waves eliminates the need for ferrite cores and as such it is possible to build a broadband linear Paul trap. Moreover, the cooling and bunching of ion beams in a square-wave-driven, three-dimensional Paul trap has been demonstrated by the Canadian Penning Trap (CPT) group at the Argonne National Laboratory (ANL)~\cite{Vaz02}. Using two fast switching, metal-oxide semiconductor field-effect transistors (MOSFETs) in a push-pull configuration, the group was able to create a square-wave with a peak-to-peak amplitude of $100 \, \textrm{V}$ at frequencies of up to $300\, \textrm{kHz}$. In addition, digital 3D ion traps have been used as mass filters at the Shimadzu research laboratory in the UK \cite{Ding:2002uq,Li-Ding:2004qy,sud02}. There, a three-dimensional Paul trap has been driven with $1\,\rm{kV}$ peak-to-peak amplitudes at up to $1\,\textrm{MHz}$. The name Digital Ion Trap (DIT) was first introduced by the Shimadzu group because digital electronics are used to control the fast switching of the applied square-wave and, in the case of the three-dimensional trap, any other applied RF excitations. 

The main limiting factor in the design of a square-wave generator is the energy dissipated in the switching transistors. This energy scales
linearly with the capacitance of the driven system. An RFQ cooler and buncher is typically much larger than any previously developed
square-wave-driven system and hence presents a larger capacitive load, e.g. $\sim$1500\,pF, for the TITAN RFQ~\cite{Smith:2006fu}. For this reason, a square-wave-driven RFQ cooler and buncher has been hitherto unrealizable experimentally. However, at TRIUMF a system has been developed for stacking MOSFETs such that the total energy dissipated in each chip was reduced. Initially developed for the MuLan experiment~\cite{1278079}, this technique was adopted in order to design a square-wave generator to drive such a trap. For TITAN's RFQ, a driver system was developed to operate at $400\,\textrm{V}_{\rm{pp}}$ switching at frequencies of up to $1 \, \textrm{MHz}$~\cite{1278079}.
\subsection{Ion motion in a Digital RFQ}
The following theoretical summary is briefly presented for the sake of completeness. A more detailed derivation of the results presented can be found in e.g.~\cite{Smi05,Smi08a}. 
The differential quations describing an ion's motion in the RFQ are so-called Meissner equations~\cite{Mei18}. They are specific versions of a more general set of equations known as the Hill equations. Hill equations have both stable and unstable solutions~\cite{Flo83} corresponding to an ion having bound or unbound trajectories in the digital RFQ, respectively. These equations can be parametrizised by two parameters $u$ and $q$:
\begin{eqnarray}\label{eq:1}
u=\frac{\omega^2}{4}x\, ,\qquad
q = \frac{4 Z e V}{m \omega^2 r_0^2}\, ,
\end{eqnarray}
with $u$ being the ion's motion with respect to just one of the radial axes and $q$ being the stability parameter. Here, $r_0$ is the radius of the ion trap (see Figure~\ref{fig:rfq}), $V$ the amplitude of the applied RF (zero-to-peak), $m$ the mass of the ion, $Z$ its charge state, and $\omega$ the angular frequency of the voltage supplied. The parameters $u$ and $q$ are identical to those usually used for the sinusoidal trap~\cite{ghosh}.

For an ideal square-wave driven RFQ with 50\% duty cycle, the stability parameter $q$ must be less than $0.712$ for the ion's motion to be stable \cite{Smi05}. In the stable region the ion motion consists of a harmonic macro motion that is slightly pertubated by a higher order micro motion. The motion is analogous to that in a sinusoidal driven trap.
%
The ion undergoes to first order a simple harmonic motion with frequency $\omega_s= \beta\, \omega/2$, which is given as~\cite{sud02}:
\begin{eqnarray}\label{eq:2.23}
\beta= \frac{1}{\pi}\:\arccos[\cos(\pi \sqrt{q / 2}) \cosh(\pi \sqrt{q / 2})]\,.
\end{eqnarray}
This leads to the idea that the ions are trapped in an harmonic time-independent pseudo-potential of depth:
\begin{equation}\label{eq:7a}
V_{pseudo} = \frac{1}{2} \frac{m \omega_s^2 u_{\rm{max}}^2}{e},
\end{equation}
with $u_{max}$ being the maximum amplitude of the macro-motion \cite{Smi05}. For low values of the stability parameter, i.e., $q < 0.3$, the depth of the pseudo-potential can be approximated as~\cite{Li-Ding:2004qy}:
\begin{equation}\label{eq:7b}
V_{pseudo} \approx \alpha q V,
\end{equation}
where $\alpha = 0.206$ for a digital trap. This same expression can be used to describe the harmonic pseudo-potential generated by a sinusoidally-driven trap in which case $\alpha = 0.125$. Hence, for a given RF amplitude a digital trap creates a pseudo-potential which is over 1.5 times deeper than a sinusoidally-driven trap.
\section{Experimental Setup}
\label{setup}
\begin{figure}
\centering
\includegraphics[width=.8\textwidth]{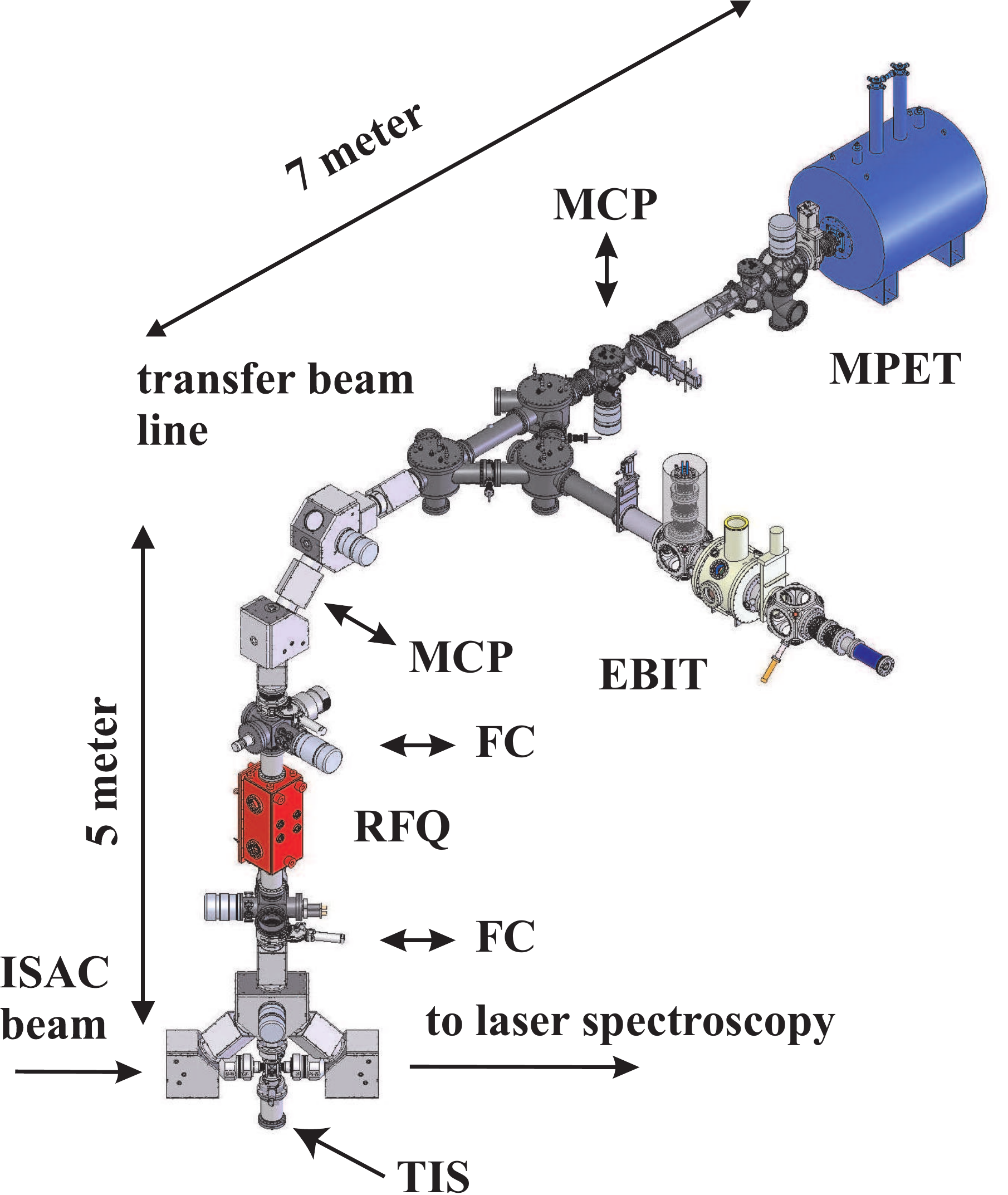}
 \caption{Schematic overview of the TITAN experiment displaying all currently installed ion traps along with the RFQ. Ion beams can be sent to TITAN either from ISAC or TITAN's test ion source (TIS).\label{fig:TITAN-setup}}
\end{figure}

\begin{figure}
\centering
\includegraphics[width=.8\textwidth]{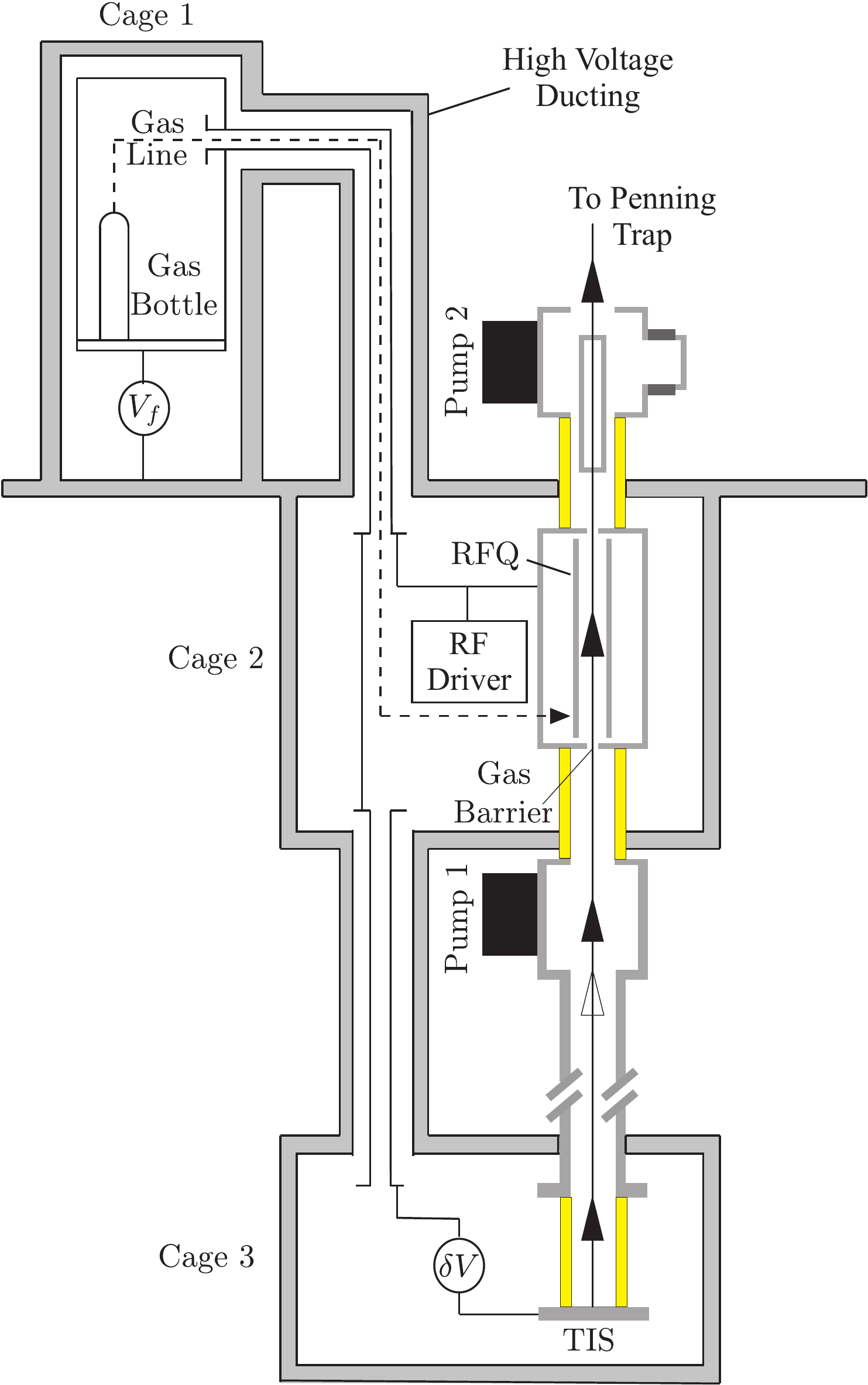}
 \caption{Schematic overview of the high-voltage set-up for the TITAN RFQ and TIS. The ceramic insulators are displayed in yellow.\label{fig:RFQSche}}
\end{figure}

The TITAN RFQ is installed in a vertical section of the beam line as the first ion trap of the TITAN experimental facility, which is located above the ISAC experimental hall ground floor on a platform. Its location is displayed in Figure~\ref{fig:TITAN-setup} along with a schematic overview of the entire TITAN experiment. The RFQ is electrically insulated from the beam line by ceramic vacuum tubes and floated at a voltage matching ISAC's ion beam energy. A Test Ion Source (TIS) located directly below the RFQ and ISAC beam line is biased at the same potential as the RFQ. It allows off-line testing of the device independent of ISAC beam. A schematic overview of the high-voltage set up is shown in Figure~\ref{fig:RFQSche} and explained in more detail in Section~\ref{sec:RFQ}. Here we provide a description of the RFQ and results from its commissioning in the TITAN setup. 

The TIS (see Figure~\ref{fig:TITAN-setup}) produces ions through thermionic emission~\cite{PhysRev.50.464} by heating a filament. So far, alkali ion sources have been used delivering stable lithium, sodium, potassium, rubidium, and cesium to TITAN's RFQ. This TIS delivers ion beams using commercial surface ion sources purchased from HeatWave Labs~\cite{Heatwave}.
An ion beam is created by floating the ion source at the same potential as the RFQ, typically between $2$ to $40\,\textrm{kV}$ above ground. In addition, the filament can be floated further ($\delta$V) in order to increase the ion's kinetic energy when entering the RFQ. The ions are extracted from the source's surface using a puller electrode biased at a few kilovolts below the source voltage. After passing this puller electrode, the ions are accelerated towards ground such that their final longitudinal kinetic energy is defined by the floating voltage of the source. The ions then pass through a series of electrostatic ion optics before being injected into the TITAN RFQ.

Independent of the ion source, i.e., TITAN's TIS or ISAC, the ions are injected into the RFQ, cooled inside the device and then extracted using specially designed optics. The following sections describe in detail these injection, RFQ, and extraction optics including the pulsed drift tube, which is used to adjust the subsequent transport energy of the ion bunch.
\subsection{Injection}
\label{sec:injec}
One of the biggest challenges in the design of a gas-filled RFQ is that of beam injection into the device. The ISAC beam transport system is designed to deliver ions to the TITAN RFQ with a transverse emittance of up to $50\, \pi \,\rm{mm} \, \rm{mrad}$ and energies ranging from $12\, \textrm{to} \, 60 \, \textrm{keV}$. The purpose of the injection optics is to accept this beam and decelerate it down to energies on the order of $10 \, \textrm{eV}$ whilst maintaining a good match between the emittance of the beam and the acceptance of the RFQ. The deceleration can be achieved by floating the RFQ at a voltage close to, but slightly below, the beam energy. The acceptance of the RFQ changes depending on the amplitude, frequency and phase of the applied RF. It is therefore usual to design the optics to match the first order acceptance of the RFQ, i.e., the harmonic ellipse corresponding to the pure macro-motion, then carry out higher-order simulations to see how well the system will accept the beam.
\begin{figure}
\centering
\includegraphics[width=0.8\textwidth]{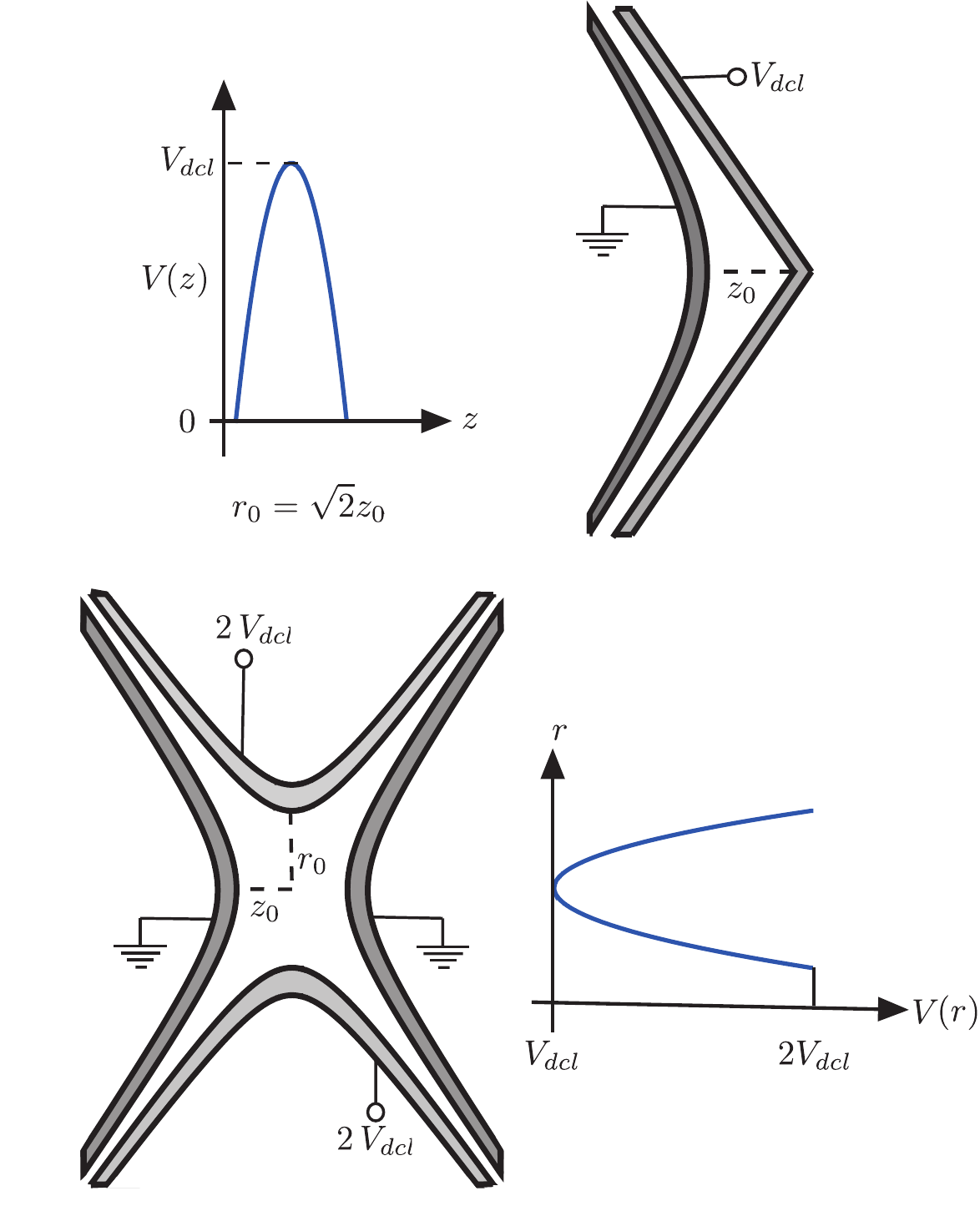}
 \caption{Schematic of two possible electrode configurations that provide a harmonic electrical potential with the required decelerating voltage $V_{dcl}$. One hyperbolic-shaped electrode in combination with one conical electrode generates this harmonic potential (top). A similar potential can be created by four hyperbolic electrodes with twice the decelerating voltage $V_{dcl}$ applied (bottom). The latter geometry is used to generate the harmonic potential in precision Penning traps. The top electrode configuration is used to decelerate the incoming ion beam. A SIMION simulation of the injection is presented in Figure~\ref{fig:decel3d}.}
 \label{fig:decel}
\end{figure}

The low-energy beam transport line was designed so that the ISAC beam's transverse emittance profile would form a right ellipse at a distance of $240 \, \textrm{mm}$ downstream from the entrance to the RFQ. For a $50 \, \pi \, \textrm{mm} \, \textrm{mrad}$ beam this right ellipse would have a radius of $2.5 \, \textrm{mm}$ in both transverse directions. It can be shown that this emittance profile overlaps well with first order acceptance of the RFQ. The deceleration optics were therefore designed so that they could decelerate the beam whilst maintaining this emittance profile and thus provide for efficient injection of the beam into the RFQ.

As previously noted, a beam undergoing simple harmonic motion traces out a right ellipse in position-velocity phase space. Therefore, the deceleration optics create a radially-harmonic potential such that the natural motion of an ion in the potential matches the initial emittance of the injected beam. Thus, the beam is decelerated without any change in its transverse emittance profile. Such a system has previously been described using an axially symmetric static quadrupole~\cite{Moore:2006lr}. However, this geometry was not suitable as it would require some re-shaping of the RFQ electrodes. Instead the design was simplified so as to only require two deceleration electrodes: one connected to ground potential and the other biased with the same voltage as the floating RFQ. The potential is generated using the electrode geometry presented in Figure~\ref{fig:decel}. As a comparison, the latter figure also displays the electrode configuration in a hyperbolic structure and the resulting harmonic potential. For the hyperbolic deceleration lens, the required radius $r_0$ can be calculated using:
\begin{equation}
\omega \propto \frac{\sqrt{E_{rad}}}{r_{beam}} = \frac{\sqrt{E_{long}}}{r_0},
\end{equation}
with $E_{rad}$ and $E_{long}$ being the initial radial and longitudinal energies of the beam and $r_{beam}$ being its radius as it enters the decelerator. This leads to the simple condition:
\begin{equation}
r_0 = \sqrt{\frac{E_{long}}{E_{rad}}}r_{beam} \approx \frac{r_{beam}}{ \theta_{beam} },
\end{equation}
where $\theta_{beam}$ is the divergence of the beam as it enters the decelerator.
The beam delivered from ISAC has $r_{beam} = 2.5 \, \textrm{mm}$ and $\theta_{beam} = 20 \, \textrm{mrad}$ so:
\begin{equation}
r_0 = 125 \, \textrm{mm}, \, z_0 = \frac{r_0}{\sqrt{2}} = 88.4 \, \textrm{mm}
\end{equation}

\begin{figure}
\centering
\includegraphics[width=1\textwidth]{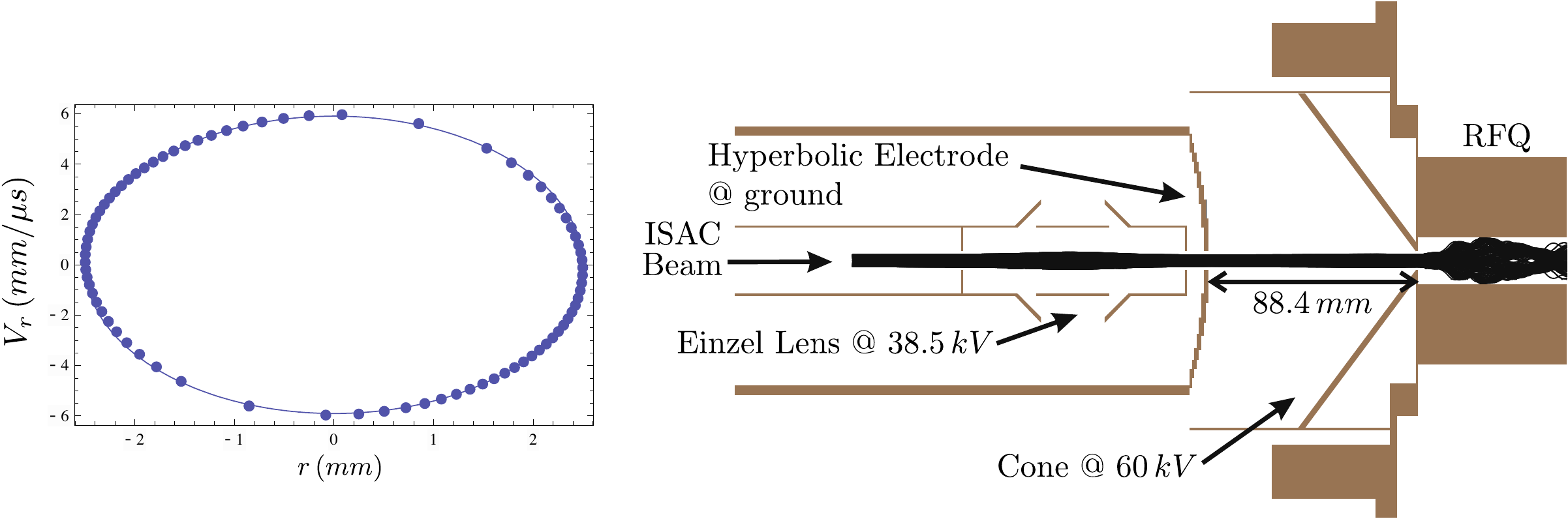}
 \caption{Comparison of the initial (solid line) and final (dots) emittance of a $60 \, \textrm{keV}$, $50 \, \pi \, \textrm{mm} \, \textrm{mrad}$ beam of mass $133\,u$ after simulated deceleration with the new injection optics (left). There are some minor differences between the two ellipses. However, they are close to identical. SIMION simulation of the new deceleration optics for the TITAN RFQ (right). Here a $40 \pi \, \textrm{mm} \, \textrm{mrad}$ beam is decelerated from $60.1 \, \textrm{keV}$ and injected into the RFQ, $q = 0.3$, $V_{pp} =400 \, \textrm{V}$, with 100\% efficiency. 
 \label{fig:decel3d}}
\end{figure}

The final design of the deceleration optics was based on simulations carried out using the ion optics package SIMION~\cite{Dahl:2000lr}. These simulations were necessary in order to take into account deviations from the ideal harmonic potential. Such deviations can be created by the introduction of apertures in the deceleration electrodes to allow for injection of the ions, and by the truncation of the electrodes which are ideally infinite. 
Two sets of simulations were carried out. The first was a simple, two-dimensional simulation of the system which ignored the effects of the electrodynamic field applied to the RFQ. This simulation was carried out on a relatively fine grid with ten grid units per millimeter. These simulations showed that although the effects of truncating the electrodes were negligible, the injection apertures act so as to slightly focus the injected beam. A simple solution to this problem was to place an einzel lens before the decelerator. The lens was placed so as to exactly compensate for the focusing effect caused by the apertures in the decelerator. Figure~\ref{fig:decel3d} (right) shows the results of the two dimensional simulation. It can be seen that with the inclusion of the einzel lens, the decelerator perturbs minimally the emittance profile of the injected beam. Finally, a full three-dimensional simulation was carried out on a coarser grid size (one grid unit per millimeter). The effects of the RF potential were included in this simulation. It was found that at an applied peak-to-peak voltage of $400 \, \textrm{V}$, a $60 \, \textrm{keV}$, $40\, \pi \, \textrm{mm} \, \textrm{mrad}$ beam could be injected into the RFQ without losses (see Figure~\ref{fig:decel3d}). However, a peak-to-peak voltage of $600 \, \textrm{V}$ is required to fully accept the $50\, \pi \, \textrm{mm} \, \textrm{mrad}$ at $60 \, \textrm{keV}$, where typical ISAC beam emittances lie between 5 and 25\,$\pi$\,mm\,mrad.
\subsection{The TITAN RFQ}\label{sec:RFQ}
\begin{figure}
 \begin{center}
 \includegraphics[viewport= 1 520 610 795, clip,width=1\textwidth]{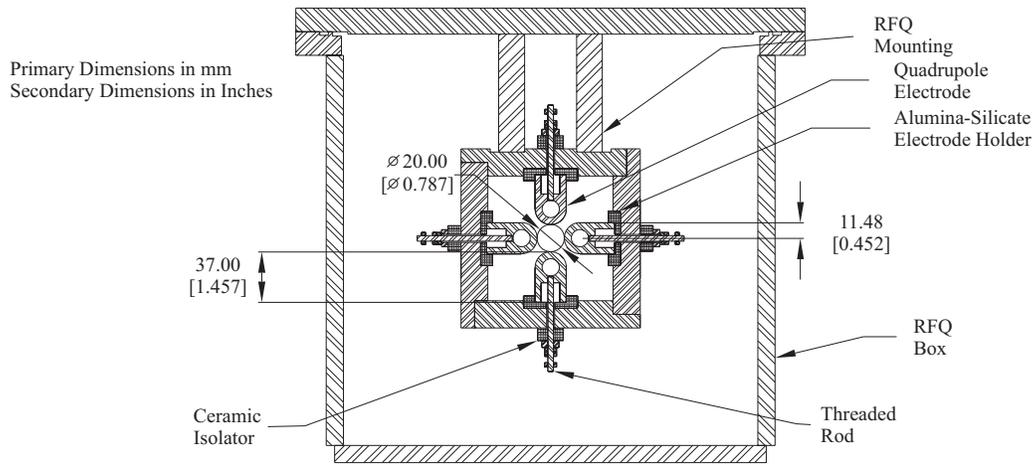}
 \caption{Mechanical drawing of the RFQ design (end view).
 \label{fig:RFQend}}
\end{center}
\end{figure}

\begin{figure}[htbp]
 \begin{center}
 \includegraphics[angle=0,width=1\textwidth]{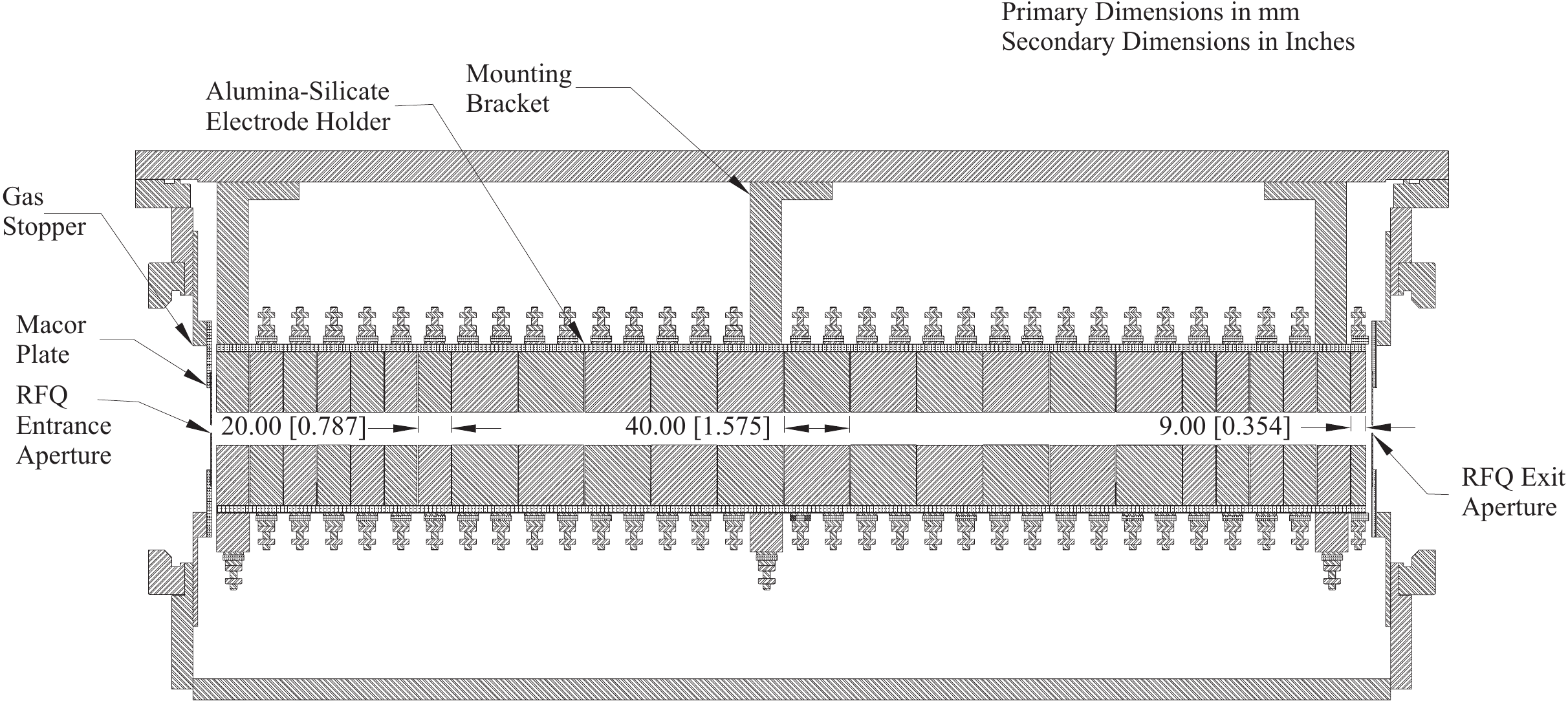}
 \caption{Mechanical drawing of the RFQ design (side view).
 \label{fig:RFQside}}
\end{center}
\end{figure}

\begin{figure}[htbp]
 \begin{center}
 \includegraphics[angle=0,width=0.51\textwidth]{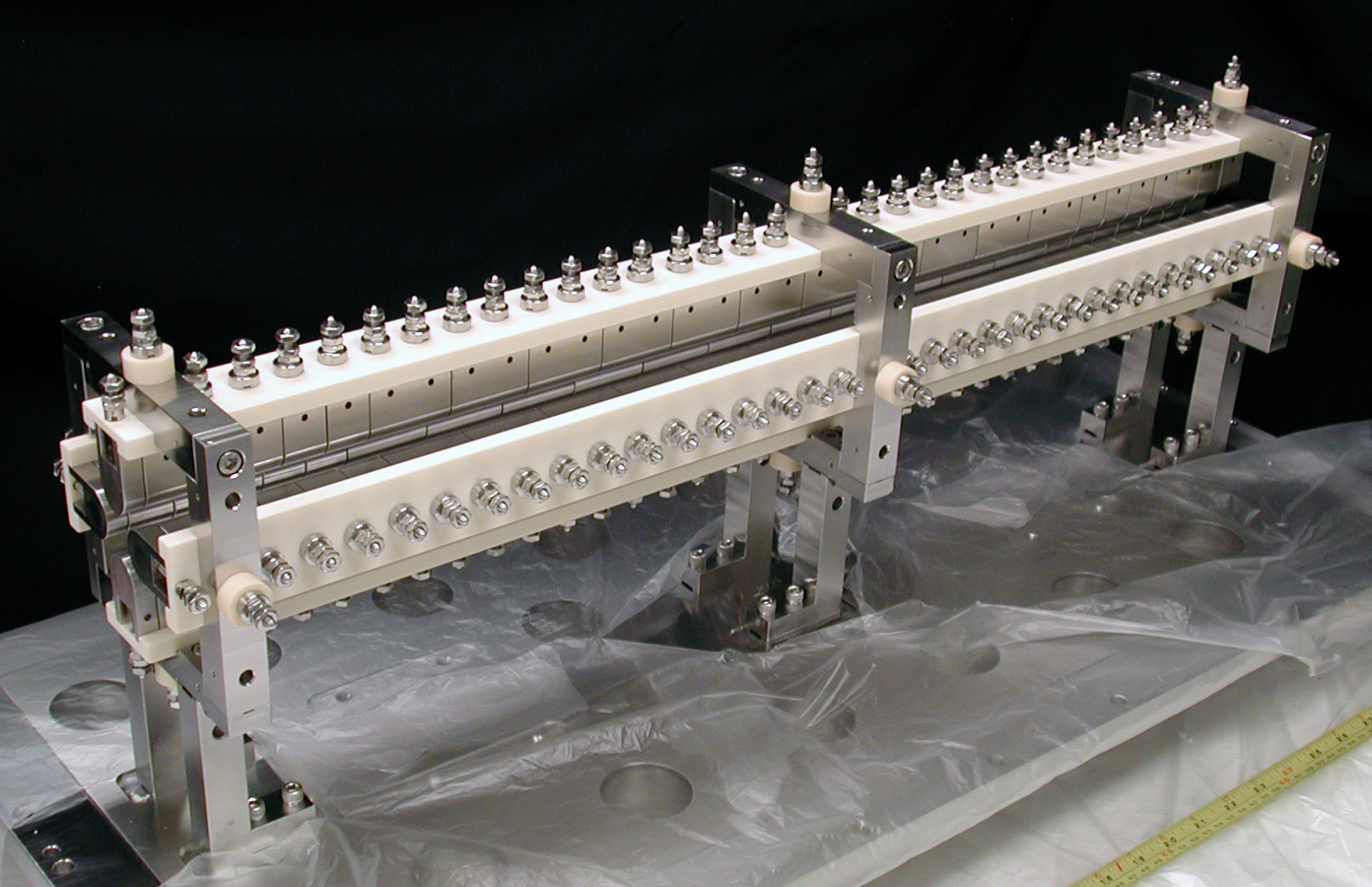}
\includegraphics[angle=0,width=0.44\textwidth]{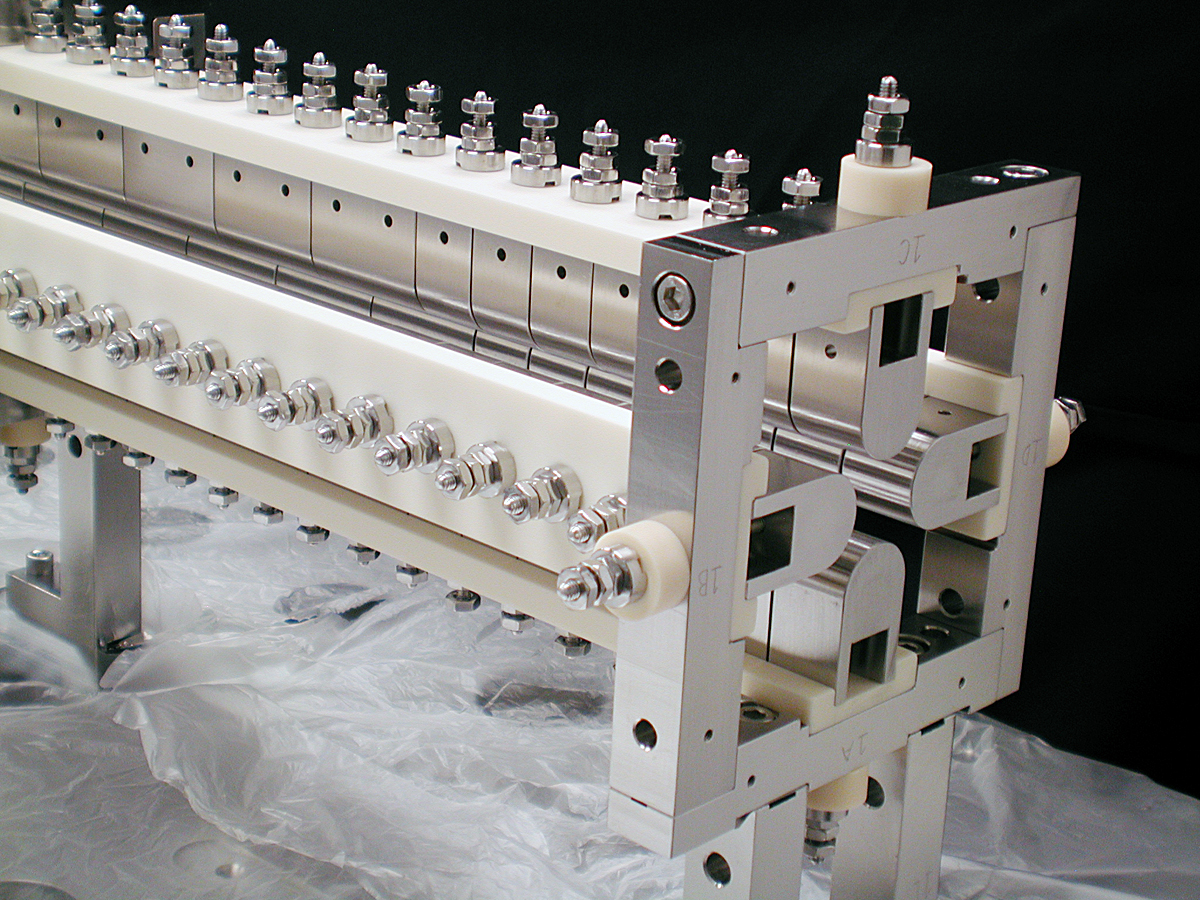}
 \caption{Picture of the electrode structure mounted on the lid of the RFQ box (left) and a close up view of the quadrupole electrode configuration (right).
 \label{fig:RFQpicture}}
\end{center}
\end{figure}

The structure of the RFQ is presented in Figures~\ref{fig:RFQend} and~\ref{fig:RFQside}. The RFQ system has an $r_0 = 10 \, \textrm{mm}$, is 
$700 \, \textrm{mm}$ long and was segmented longitudinally into twenty four pieces. The 
main body of the trap consists of eleven segments, each $40\, \textrm{mm}$ long. Twelve 
$20\, \textrm{mm}$ sections were used at the ends of the trap, seven at the trap entrance 
and five at the trap exit, along with one $9\, \textrm{mm}$ piece. The shorter electrodes 
were used so as to give greater control over the longitudinal potential in the 
injection and bunching regions. A picture of the electrode structure is presented in Figure~\ref{fig:RFQpicture}.

The RFQ electrodes were machined in a `pillbox' shape. This allowed
for the electrodes to be easily mounted on four precision-machined,
alumina-silicate electrical isolators. These isolators were designed
to hold the electrode pieces such that they were equally spaced
and properly aligned. The isolators were held in place by three
metal frames mounted on the lid of the RFQ box (see Figure \ref{fig:RFQside}). This allowed one to easily access the trap structure for maintenance work. Opposing pairs of electrodes were wired together using stainless
steel rods. The box lid had fifteen $2\frac{3}{4}''$
ConFlat flanges welded to it to allow for the mounting of
electrical feedthroughs and the attachment of a gas feeding
system. 

In order to decelerate the incoming beam prior to cooling, the RFQ is floated at a voltage, $V_f$, close to that of the beam energy. To achieve this, the vacuum vessel housing the RFQ-electrode structure is insulated from ground potential, i.e., the beam line, using ceramic insulators capable of holding $60 \, \textrm{kV}$ electrostatic potential. The insulators' positions are indicated by yellow bars in Figure~\ref{fig:RFQSche}.
The whole device was mounted inside a Faraday cage (cage 2 in Figure~\ref{fig:RFQSche}). The square-wave driver was also placed inside the cage on ceramic insulators, and floated to the same potential as the RFQ. A second Faraday cage was built on the platform and electrically connected to the first (see schematic Figure~\ref{fig:RFQSche}); all the electronics required for the RFQ were placed within. The gas feeding system for the RFQ was also placed inside this cage. The two cages were floated at a common potential and connected via a duct with cables passing from one cage into the other. The TIS (see Figure~\ref{fig:TITAN-setup}) was placed inside a third Faraday cage connected through a second duct to the RFQ cage. Therefore, the source was floated using the same supply as the RFQ, though an additional supply was used to apply a small offset in voltage, $\delta V$, between the ion source and RFQ. This offset defined the injection energy of the ion beam as it entered the RFQ. 
\subsection{Extraction of Cooled Bunches}
\begin{figure}
\centering
\includegraphics[width=1\textwidth]{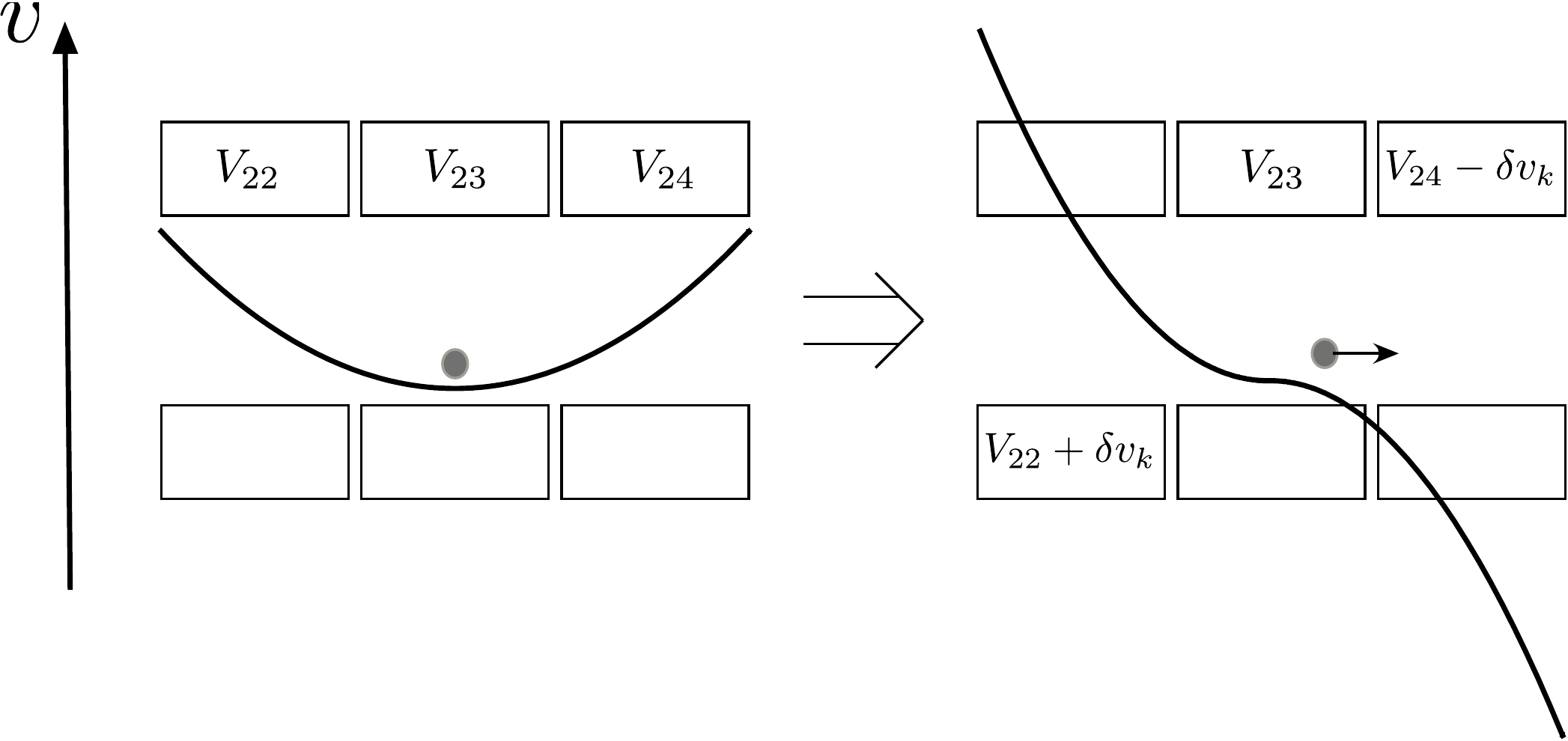}
 \caption{Extraction of an ion pulse from the RFQ is achieved by applying a kicking voltage $\delta V_k$ to the last (\#24) and third from last (\#22) segments.\label{fig:kick}}
\end{figure}

\begin{figure}
\centering
\includegraphics[width=1\textwidth]{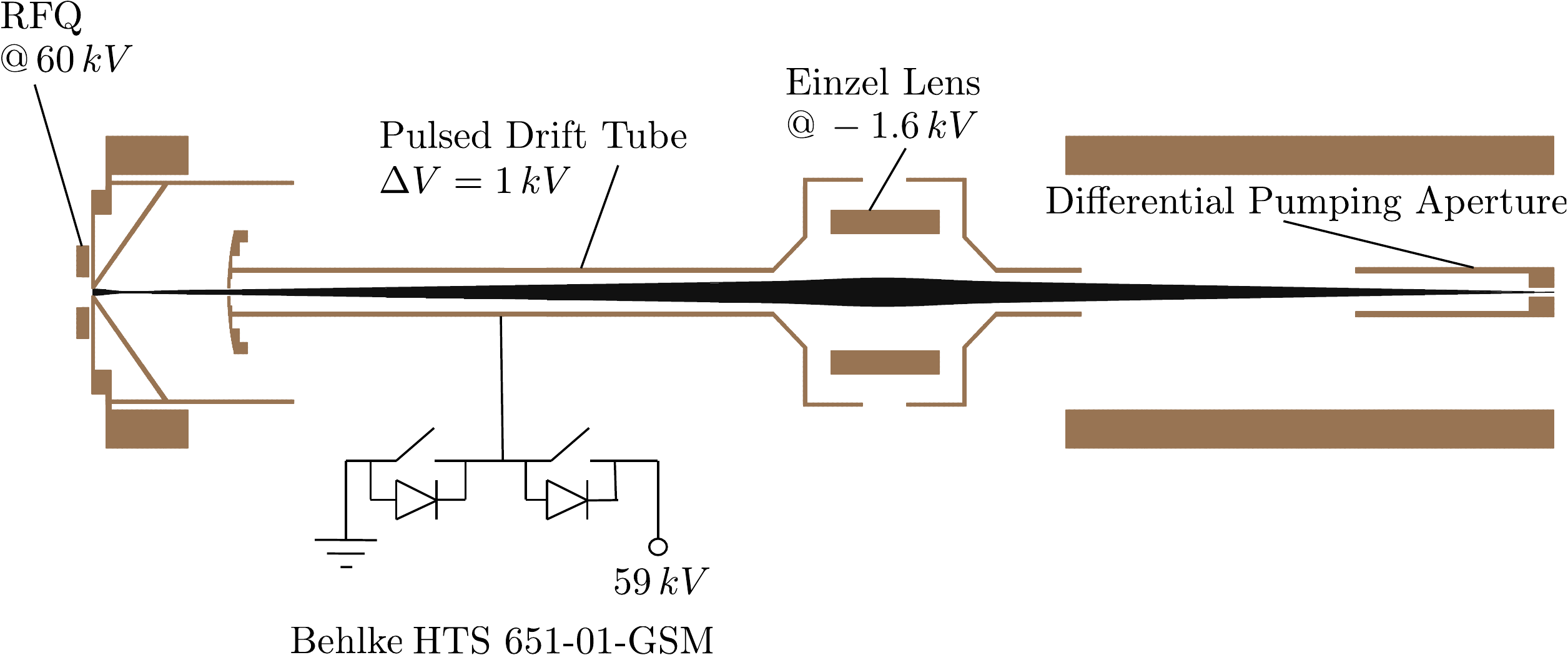}
 \caption{Extraction optics of the TITAN RFQ. An ion pulse accelerated to $1\, \textrm{keV}$ longitudinal kinetic energy before entering a pulsed drift tube. The potential on the tube is then switched using a push-pull switch such that the ions leave the tube at ground. The ions are subsequently focused so that they pass through a $5\, \textrm{mm}$ differential pumping aperture using an einzel lens placed half way between the RFQ and the aperture.
 \label{fig:ext}}
\end{figure}

Extraction of an ion pulse from the RFQ is achieved by pulsing the DC potentials applied to the RFQ electrodes as shown in Figure~\ref{fig:kick}. The ions are then re-accelerated using electrostatic optics.
\label{sec:reacc}
A SIMION-simulated rendering of the extraction optics including a simulated ion beam is illustrated in Figure~\ref{fig:ext}. The ions are extracted from the RFQ and accelerated using the inverse of the deceleration optics described in Section~\ref{sec:injec}. They are then focused using an accelerating einzel lens and passed through a differential pumping barrier. In continuous mode (DC) of operation, the longitudinal energy of the extracted beam is equal to that of the injected energy reduced by $\delta$V as the ions are re-accelerated towards ground. However, in pulsed mode (AC) the transport energy $\Delta$V of the extracted bunch can be defined by accelerating the ions into a drift tube (similar to the concept used at ISOLTRAP~\cite{Her01}) at a potential $V_f-\Delta$V as presented in Figure~\ref{fig:ext}. Once the ions are inside the tube, its potential can be switched to ground. Thus, the ions leave the tube at ground potential with a kinetic energy equal to the initial potential difference between the RFQ and drift tube, $\Delta V$. Following the drift tube, the beam is bent by two $45^{\circ}$ electrostatic benders before it is delivered vertically to either the measurement Penning trap \cite{Bro11} or electron beam ion trap \cite{Lap10} (see Figure~\ref{fig:TITAN-setup}). The twenty four RFQ electrodes, including injection and extraction optics, are illustrated in Figure~\ref{fig:RFQ-RE}. This figure also displays the drag potential applied in forward extraction (top schematic).
\subsection{Reverse Extraction of Cooled Bunches}
\begin{figure}
\centering
\includegraphics[width=1.0\textwidth]{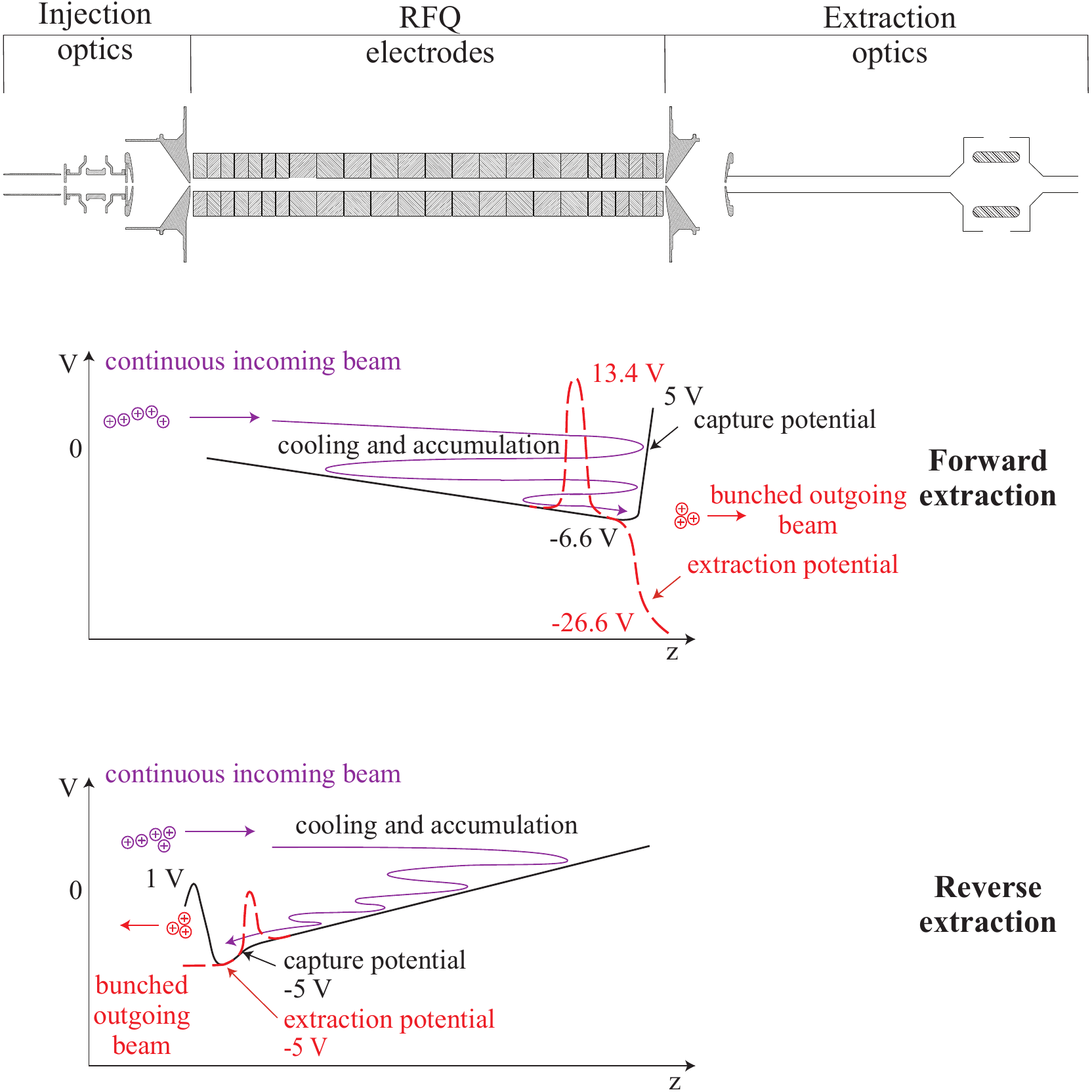}
 \caption{Schematic of the RFQ electrode configuration along with the electrical drag potential applied to the RFQ electrodes during forward (top potential) and reverse extraction (bottom potential) operation. During injection the potential presented by the solid line is applied, while for ion bunch extraction the dashed potential is applied. Typical values for the potential depths are also presented.\label{fig:RFQ-RE}}
\end{figure}

The symmetric arrangement of electrodes inside the TITAN RFQ (see Figure~\ref{fig:RFQside}) allows one to apply a mirror drag potential to the RFQ electrodes. In this case, ions are cooled in collisions with the buffer gas but instead of being dragged through the RFQ in forward direction, they are collected at the potential minimum at the entrance of the RFQ. Figure~\ref{fig:RFQ-RE} (bottom schematic) displays the applied drag potential for reverse extraction and as a comparison, also the field applied during forward extraction. This cooled ion bunch can then be extracted towards the ISAC beam line. Following extraction, the ions are accelerated towards the ground potential of the beam line since no pulsed drift tube is installed at the entrance side of the RFQ. Passing two $45^{\circ}$ electrostatic benders, the bunched and cooled ion pulse is then sent towards the laser spectroscopy beam line. So far, several stable and radioactive isotopes (e.g. $^{6,7}$Li, $^{23}$Na, $^{74-76,78,85,87}$Rb,$^{206,208}$Fr) have been extracted from the RFQ in reverse direction. A detailed description of the on-line laser spectroscopy experiment with bunched ion beams at TRIUMF is presented in \cite{Man10,Man11}. The transmission efficiency in reverse extraction will be investigated in future measurements.
\section{Systematic studies with TITAN's RFQ}
\begin{figure}
\centering
\includegraphics[width=1\textwidth]{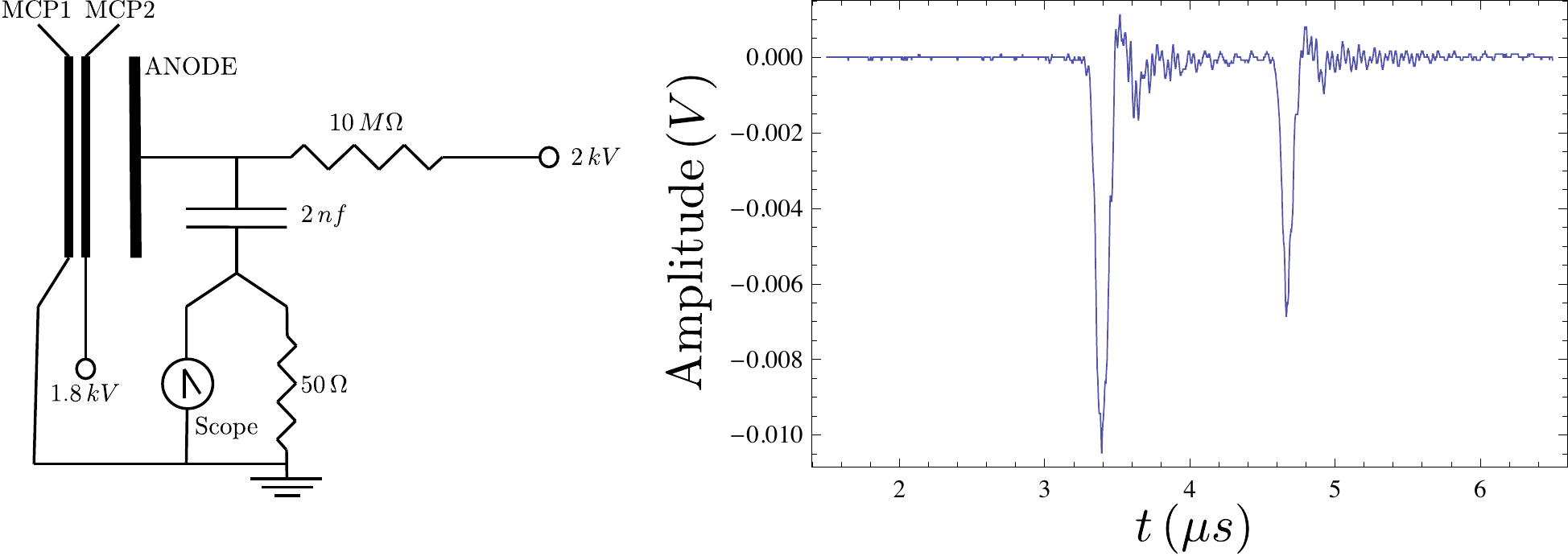}
 \caption{Circuit diagram of the connector box where the MCP signals were capacitively coupled into an oscilloscope (left). The resulting signals induced on the MCP corresponding to $^6$Li and $^7$Li are displayed on the right.\label{fig:MCPw}}
\end{figure}

During the commissioning and first experiments with radioactive isotopes, systematic studies with the TITAN RFQ were performed. For these studies, several micro-channel plates (MCPs) \cite{Wiz79} were used along the TITAN beam line (for position see Figure~\ref{fig:TITAN-setup}) as well as Faraday cups (FC) positioned in front of and behind the RFQ. Typically, the MCPs were used for single ion counting and beam optics optimization in bunched operation mode, while the FCs were used to read the ion current of a continuous beam. The MCPs were operated with the front plate at ground potential. The amplified ion signal of the MCP was then read from the anode by capacitively decoupling the signal. A typical accumulated signal of 128 individual ion pulses impinging on the MCP, along with the circuit used to decouple the signal, is displayed in Figure~\ref{fig:MCPw}.
%
\subsection{Transmission Efficiencies}
Typically, only one ion at a time is required for mass measurements performed in a Penning trap. However, when measuring rare radioactive isotopes production, yields can be as low as a few ions per second with half lives below 10\,ms, thus requiring an efficient beam transportation system. Therefore, the TITAN RFQ was designed and optimized to cool ions in a short period of time and operate with high transmission efficiencies. The transmission efficiency was investigated in both DC and bunched modes.
\subsubsection{Continuous Operation}
The transmission efficiency of the RFQ in continuous mode was studied systematically as a function of gas pressure, RF voltage, RF frequency and injection energy ($\delta V$). The DC efficiency was defined as the ratio of the extracted ion current, as measured on an FC placed after the RFQ, to the injected current measured on an FC in front of the RFQ. The tests were carried out for both cesium and lithium, at initial beam energies of $5 \, \textrm{keV}$ and $20 \, \textrm{keV}$, respectively. The emittance of the test ion source beam was limited by the geometry of the ion source system and was predicted, through SIMION simulations, to be on the order of $\epsilon_{99 \%} = 30 \, \pi \, \textrm{mm} \, \textrm{mrad}$. An initial ion current on the order of a nano-ampere was used, and a $7 \, \textrm{V}$ DC gradient was applied across the length of the RFQ.

\begin{figure}
\centering
\includegraphics[width=1\textwidth]{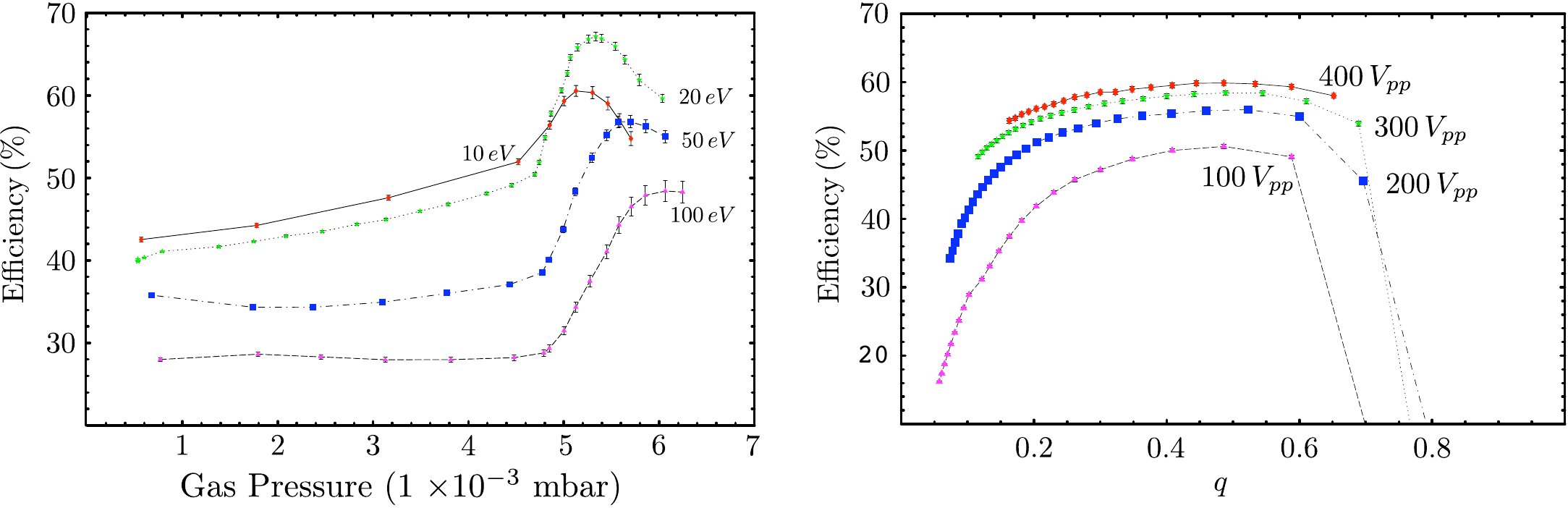}
 \caption{ The DC transmission efficiency of cesium ions in helium through the TITAN RFQ for different injection energies at various gas pressures (left) and the transmission efficiency for different $V_{pp}$ as a function of the stability parameter $q$ for a constant pressure $p = 5 \times 10^{-3} \, \textrm{mbar}$. \label{fig:cseff}}
\end{figure}

Figure~\ref{fig:cseff} shows the transfer efficiency of the cesium in helium as a function of the buffer gas pressure for different injection energies ($V_{pp} = 400 \, \textrm{V}, \, f= 600 \, \textrm{kHz}$), as well as the efficiency as a function of $V_{pp}$. In general, the efficiency of the RFQ increases as the initial energy of the ions decreases. This is because the ions are more likely to be scattered into the RFQ electrodes through high energy collisions than low energy collisions. At very low energies the efficiency falls off, probably due to the longitudinal energy spread of the beam. In this energy regime a portion of the beam will not be sufficiently energetic to enter the cooler. The efficiency as a function of the stability parameter $q$ is also shown in Figure~\ref{fig:cseff} ($\delta V = 20 \, \textrm{V}, \, p = 5 \times 10^{-3} \, \textrm{mbar}$). The maximum transmission occurs at $q \approx 0.5$. The advantage of running at higher voltages is evident as the efficiency rises with the peak-to-peak voltage.

\begin{figure}
\centering
\includegraphics[width=1\textwidth]{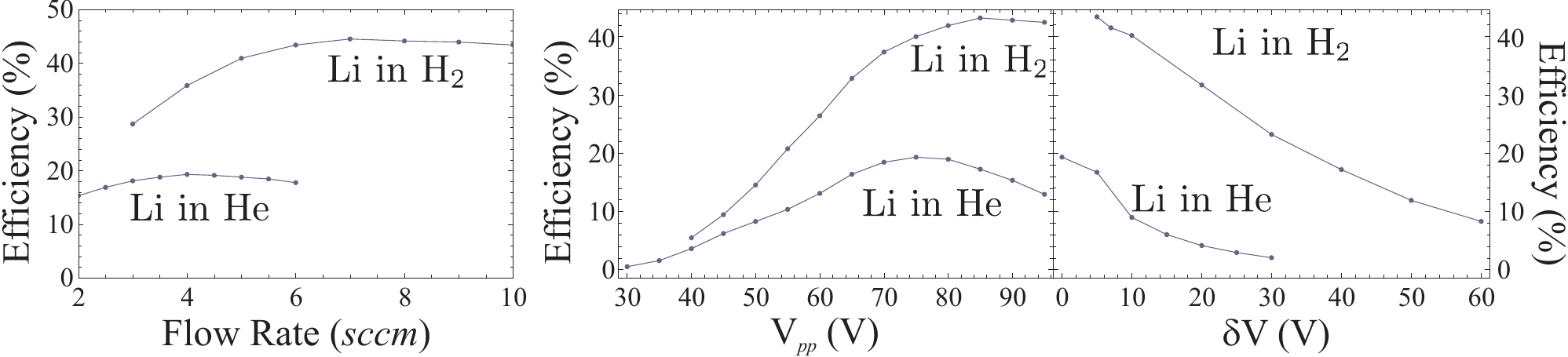}
 \caption{ The DC transmission efficiency of lithium ions in helium and H$_{2}$ through the TITAN RFQ as a function of flow rate (left), RFQ square-wave-voltage $\textrm{V}_{pp}$ (middle), and offset voltage $\delta V$ (right). For explanation see text.\label{fig:lieff}}
\end{figure}

Figure~\ref{fig:lieff} displays the DC efficiency of lithium in both molecular hydrogen and helium as a function of the flow rate of the gas into the RFQ ($f = 1.15 \, \textrm{MHz}$, $V_{pp} = 80 \, \textrm{V}$). It can be observed that with both gases, good DC efficiencies can be obtained although the peak DC efficiency in hydrogen ($\approx 40 \%$) is twice that of helium ($\approx 20 \%$). Due to the RF driver the frequency could not be increased above $1.2 \, \textrm{MHz}$. This limited the RFQ to be operated below the maximum of $400 \, \textrm{V}_{pp}$ but still provided a suitable trapping parameter, $q$, for lithium ions. The voltage of the applied RF was thus varied at a fixed frequency ($f = 1.15 \, \textrm{MHz}$) to determine the peak transfer efficiency. This occurred at slightly different voltages for the two gases but was generally higher than that for operation of the RFQ without gas. Also of note is the efficiency as a function of $\delta V$ (as defined in Figure~\ref{fig:RFQSche}). Unlike for cesium, in helium gas a lower $\delta V$ always resulted in an increased transfer efficiency. Due to the low mass of lithium, the RFQ had to be run at a lower peak-to-peak voltage than for cesium resulting in a shallower pseudo-potential (see Eq.~(\ref{eq:7a})). This, combined with a large scattering angle associated with collisions between ions and gas atoms that are similar in mass, leads to significant losses for higher energy collisions. All these studies were performed using TITAN's TIS that is expected to provide an ion beam with lower emittance than ISAC. However, during several online experiments using ISAC beam DC transmission efficiencies of more than 70\% have been achieved. They are comparable with those achieved with the TIS. 
\subsubsection{Pulsed Operation}
During an on-line experiment with radioactive $^{124}$Cs and $^{126}$Cs (t$_{1/2}=30.8(5)$\,s and $1.64(2)$\,min, respectively \cite{NNDC}) \cite{Bru10}, the transmission through the RFQ in bunched mode was investigated. Two sources of cesium were used: on-line produced at ISAC and stable from the TITAN TIS (see Figure~\ref{fig:TITAN-setup}). Radioactive cesium was produced by bombarding a Ta target with a 50\,\textmu A proton beam of 500\,MeV. Cesium ions were produced by surface ionization and delivered to TITAN with a beam energy of 15\,keV. There, they were cooled and bunched by the TITAN RFQ and sent to a $\beta$-particle counter. The counter consisted of an aluminum foil placed in front of a Si surface barrier detector. Ions were implanted into the foil where $\beta$ particles from radioactive decays impinged on the Si detector and were counted. This setup and its position along the TITAN beam line are described in \cite{Bru08}. The total detection efficiency for $\beta$ particles in this decay is estimated to be $(25\pm10)$\,\%. This efficiency
is the product of intrinsic detection efficiency and geometric detection efficiency. The latter is the main contribution to the uncertainty of the total efficiency due to the distance of $(6\pm1)$\,mm between detector and Al foil. The intrinsic detection efficiency
was calculated by integrating the $\beta$ spectrum \cite{Krane}. Based on this, an intrinsic detection efficiency of $\sim99\%$ was determined assuming that electrons below 250 keV were stopped by the 20\,\textmu m thick Al foil and the detector's dead layer, and thus not detected.

The ion bunch intensity was determined by implanting 10 ion bunches at 10\,Hz repetition rate into the Al foil. Counting with a scaler began 1006.7\,ms after the first bunch was extracted from the RFQ, i.e., directly after all bunches hit the Al foil. The flight time for $^{126}$Cs$^{1+}$ from the RFQ to the $\beta$ counter with 2\,keV transport energy was about 120\,\textmu s. The decay curve was then fitted with an exponential decay providing the half life and measured background count rate as fixed input parameters. Based on this analysis, the beam intensity was determined to be $2.8(11)\cdot10^5$ and $4.4(21)\cdot 10^5$ ions per bunch at an extraction rate of 10\,Hz for $^{124}$Cs and $^{126}$Cs, respectively. The ion beam current entering the RFQ was measured with a FC to be $I_{\textnormal{ISAC}}\left(^{124}\textnormal{Cs}\right) \sim 3 \cdot 10^{-12}$\,A $\approx 19 \cdot 10^6$ ions/s and $I_{\textnormal{ISAC}}\left(^{126}\textnormal{Cs}\right) \sim1\cdot10^{-11}$\,A$\approx 6\cdot 10^7$ ions/s. The uncertainty of the Faraday cup reading was estimated to be $0.1\cdot10^{-11}$\,A. Based on these measurements,
the efficiency of the RFQ was estimated to be about 7-15\% in bunched operation at an extraction rate of 10\,Hz. The obtained values are presented in Table~\ref{tab:transmission}. However, it was observed in later studies that the applied gas pressure was not sufficient to stop all incoming ions inside the RFQ. Thus, the achievable transmission efficiency might even be higher than the determined 7-15\%. Furthermore, if the space-charge limit of the RFQ was reached with the incoming beam this would also result in a reduced determined transmission efficiency. Similar transmission efficiencies in bunched mode using the TIS were achieved in preliminary measurements with a nuclear pre-amplifier. Further studies are planned to determine the space-charge limit of the TITAN RFQ.
\begin{table}
\centering
 \begin{tabular}{c c c c c}
\hline
Isotope&Intensity per&Intensity&Transmission\\
&bunch at 10\,Hz&at FC [A]&[\%]\\
\hline
$^{124\textnormal{g}}$Cs&2.8(11)$\cdot 10^5$&$\sim 3 \cdot 10^{-12}$&$15(6)$\\
$^{126}$Cs&$4.4(21)\cdot10^5$&$\sim1\cdot 10^{-11}$&$7(3)$\\
\hline
 \end{tabular}
\caption{$^{124,126}$Cs ion bunch intensities determined while operating the RFQ with an extraction rate of 10\,Hz. The injection parameters were different for $^{124}$Cs and $^{126}$Cs. This could explain the different transmission efficiencies.}
\label{tab:transmission}
\end{table}

In the case of $^{124}$Cs, a 2\% contamination of $^{124\textnormal{m}}$Cs was present and considered in the analysis. During the measurement with $^{126}$Cs, no measurable amount of isobaric contamination was present.
\subsection{Life times in the trap and cooling time}
\begin{figure}
 \centering
\includegraphics[width=.6\textwidth]{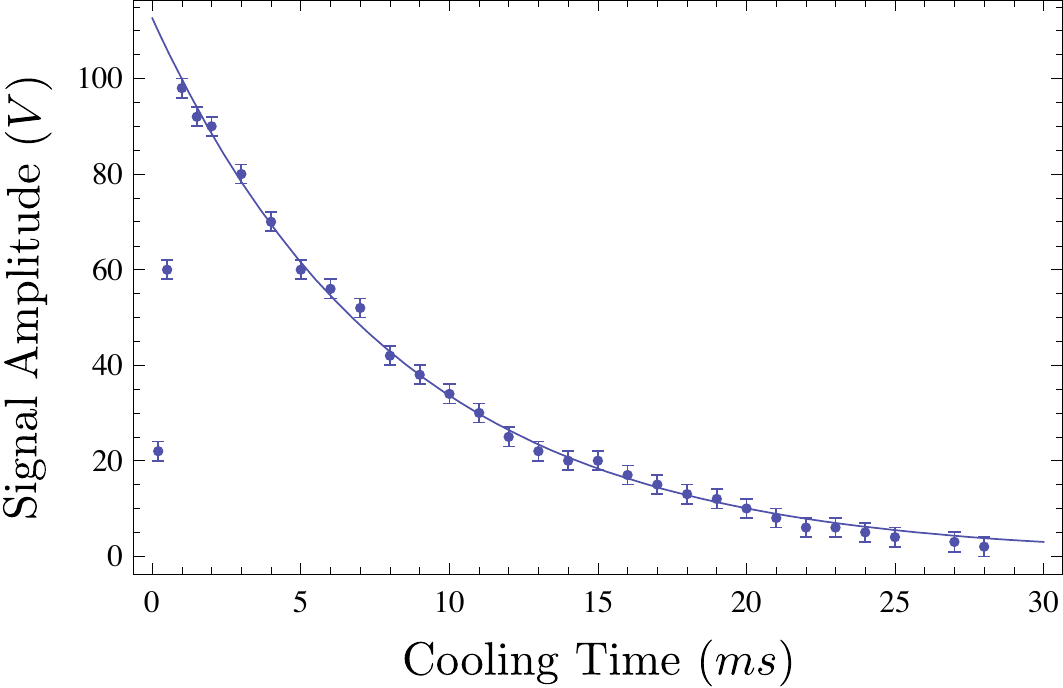}
\caption{Signal amplitude of lithium ions impinging onto an MCP detector as a function of cooling time in He buffer gas at $p\approx 4.5 \times 10^{-3} \, \textrm{mbar} $.\label{fig:hl-li}}
\end{figure}

The cooling of an ion through elastic scattering is accompanied by charge exchange reactions that result in neutralization and hence, the ion's loss inside the trap. Among other factors, this exchange depends on the chemical properties of the injected ions as well as the applied buffer gas and contained impurities. An ion gate was used in front of the RFQ to only inject ions for a short period of time, typically 100\,\textmu s. Following injection, the ions were cooled for a certain time period and then extracted onto an MCP detector. A multi-channel scaler was then used to record the ion's time-of-flight along with the total number of ions extracted. This measurement was performed using lithium and cesium ions. For lithium, the life times in He and H$_2$ were investigated while cesium was cooled using helium. For lithium in H$_2$ and cesium in He, no appreciable difference in the number of extracted ions could be observed for cooling times up to 30\,ms. However, a clear decay was observed if lithium was cooled using He as coolant. In this case, a half life of $5.7\pm0.1$\,ms was observed as displayed in Figure~\ref{fig:hl-li}. In the case of Li with He being the coolant RF heating \cite{PhysRev.170.91} increases the kinetic energy of the Li ion and thus reduces its life time inside the cooler.
%
\subsection{Pulse Widths}
\begin{figure}
\centering
\includegraphics[width=.75\textwidth]{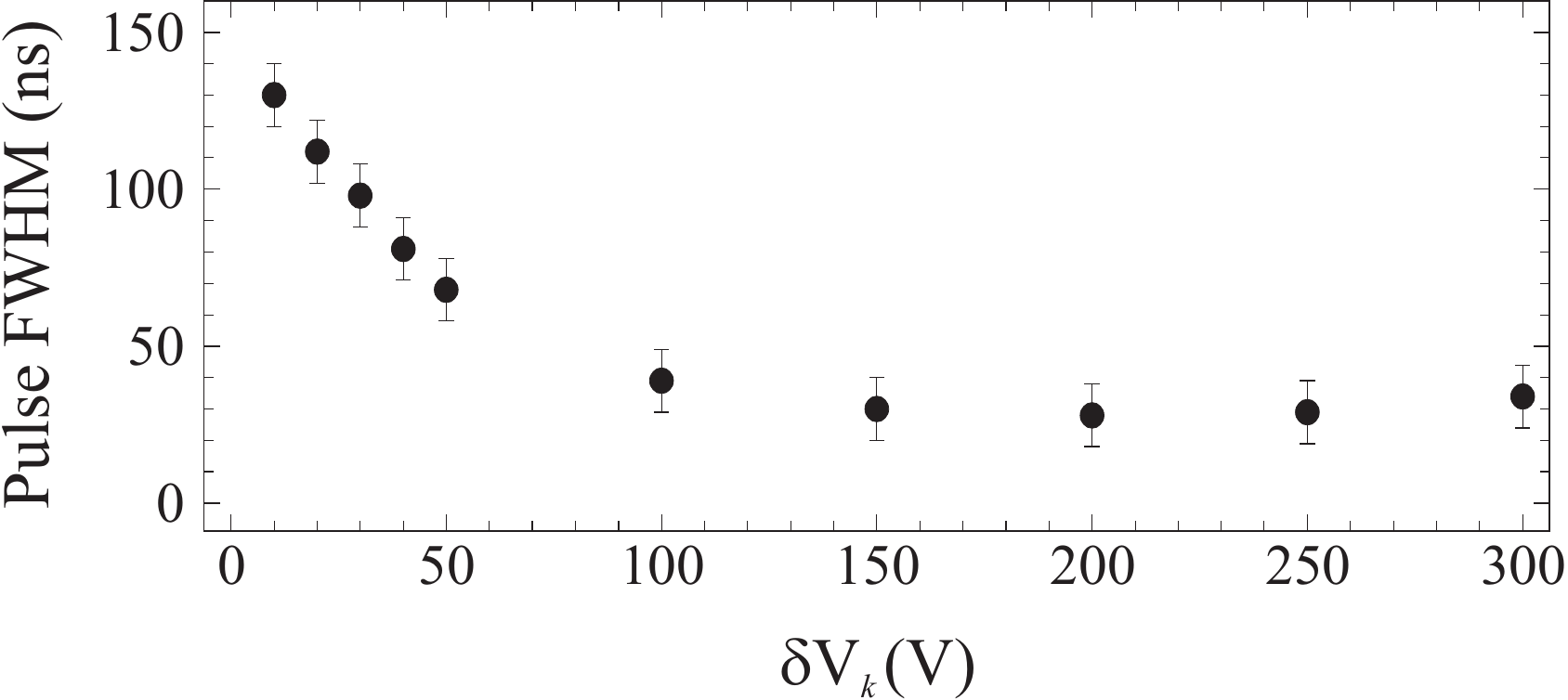}
 \caption{Measured pulse widths (Full Width Half Maximum) as a function of extraction voltage, $\delta_{V_k}$ without using the pulsed drift tube. It can be seen that without the pulse drift tube, pulse widths on the order of $30 \, \textrm{ns}$ were achievable. 
\label{fig:pw}
}
\end{figure}

Pulsed operation of the RFQ was demonstrated by detecting the extracted pulses on an MCP detector. The signal induced on the MCP's anode was capacitively coupled onto an oscilloscope as shown in Figure~\ref{fig:MCPw}. The ions were collected by keeping the same $7 \, \textrm{V}$ DC potential, but with the potential on the last RFQ electrode raised above the high voltage bias. For cesium ions, a deep well was formed by lowering the potential on electrode (23) to $-11 \, \textrm{V}$. However, for lithium ions it was observed that this would result in ion pulses of severely reduced amplitude. A typical pair of detected lithium pulses is shown in Figure~\ref{fig:MCPw}. There are two peaks; the earliest and therefore faster corresponds to the lighter isotope $^{6}$Li, the latter to $^7$Li. The width of the ion pulses as a function of the extraction voltage is shown in Figure~\ref{fig:pw}. It can be seen that extremely short pulses ($30 \, \textrm{ns}$ FWHM) were achievable. The pulse width was observed to decrease with increased extraction amplitude, with a plateau at around $100 \, \textrm{V}$. 
\section{Conclusion and future developments}
The RFQ presented here is the first ion trap of the TITAN experiment. It is designed to cool and bunch radioactive ion beams delivered from ISAC on a short time scale with high efficiency, and has been successfully installed and commissioned at the TITAN setup. It is operated as a square wave driven system which allows broad band application of high RF amplitudes for maximum transfer efficiencies for all elements ranging from He to Fr (mass 4\,u to 220\,u). The RFQ has been used to decelerate, cool and bunch the incoming continuous ion beam in off-line studies as well as during several experiments with radioactive ions. Transmission efficiencies of 70\% were achieved in DC operation mode, while up to 7-15\% were reached in bunched mode with $\sim 10^5$ ions per bunch at a repetition rate of 10\,Hz. These transmissions are comparable to efficiencies of RFQs installed at other isotope facilities.
The transmission of light lithium ions through the RFQ was investigated for He and H$_2$ as buffer mediums. It was found that the transmission of $^6$Li is increased by a factor of two in H$_2$ compared to He. Furthermore, the use of H$_2$ made it possible to extract radioactive $^{8}$He out of the RFQ. This allowed for the first Penning trap measurement of this isotope~\cite{Ryj08}. 

The unique quasi-symmetric design of the trap electrodes enables operation of the RFQ in reverse extraction mode. The possibility of delivering cooled ion bunches from TITAN's RFQ to other ISAC experiments makes this device a versatile tool for a wide range of high precision experiments at low beam energies; in particular, for laser spectroscopy. Systematic studies of the transmission efficiency in reverse extraction mode are planned to compare them to those in forward direction.
%
\section{Acknowledgements}
We would like to dedicate this publication to the inventor of RFQ devices at radioactive ion beam facilities, Prof. R.B. Moore, who recently passed away.

TRIUMF receives federal funding via a contribution agreement with the National Research
Council of Canada (NRC). This work was funded by NSERC and NRC. One of the authors (TB) acknowledges
support by evangelisches Studienwerk e.V. Villigst. A.G. acknowledges support from NSERC
PGS-M program and S.E. from the Vanier CGS program. The pictures displayed in Figure~\ref{fig:RFQpicture} were provided by TRIUMF.


\bibliographystyle{elsarticle-num}
\bibliography{bibliography}

\end{document}